\documentclass{ws-ijmpe}
\newcommand{\tauiso}{{\mbox{\boldmath $\tau$}}}
\newcommand{\gmu}{{\gamma_\mu}}

\begin{document}


\catchline{}{}{}{}{}

\title{OPEN AND HIDDEN STRANGENESS PRODUCTION IN NUCLEON-NUCLEON  
COLLISIONS\footnote{Lecture presented in the 4th DAE-BRNS workshop on hadron
physics, Aligarh Muslim University, Aligarh, India, Feb. 18-23, 2008}}

\author{\footnotesize Radhey Shyam }

\address{Theory Division, Saha Institute of Nuclear Physics\\
1/AF Bidhan Nagar, Kolkata 700064, India \\
radhey.shyam@saha.ac.in}

\maketitle

\begin{history}
\received{(received date)}
\revised{(revised date)}
\end{history}

\begin{abstract}

We present an overview of the description of $K$ and $\eta$ meson productions
in nucleon-nucleon collisions within an effective Lagrangian model where meson
production proceeds via excitation, propagation and subsequent decay of 
intermediate baryonic resonant states. The $K$ meson contains a strange quark
($s$) or antiquark ($\bar s$) while the $\eta$ meson has hidden strangeness as
it contains some component of the $s{\bar s}$ pair. Strange meson production is
expected to provide information on the manifestation of quantum chromodynamics
in the non-perturbative regime of energies larger than that of the low energy
pion physics. We discuss specific examples where proper understanding of the 
experimental data for these reactions is still lacking.   
 
\end{abstract}

\section{Introduction}

It is well established that nucleons have a rich excitation spectrum which
reflects their complicated multi-quark inner dynamics. The determination of
properties of the nucleon resonances (e.g., their masses, widths, and coupling
constants to various decay channels) is an important issue of the hadron
physics. This will provide the benchmark for testing the predictions of the
Lattice quantum chromodynamics (LQCD) which is the only theory which tries to
calculate these properties from the first principles~\cite{wil74}. Even though,
the requirement of the computational power is enormous for their numerical
realization, such calculations have started to provide results for properties
of ground as well as excited states of the nucleon~\cite{bur06,lei05}.
Furthermore, reliable nucleon resonance data are also important for testing the
"quantum chromodynamics (QCD) based" quark models of the nucleon (see, e.g.
\cite{cap00,lor01}) and also the dynamical coupled-channels models of baryonic
resonances~\cite{lut05}.

Experimental determination of baryonic resonance properties proceeds indirectly
by exciting the nucleon with the help of a hadronic or electromagnetic probe
and performing measurements of their decay products (mesons and nucleons). The
reliable extraction of nucleon resonance properties from such experiments is a
major challenge. As description of the intermediate energy scattering is still
far away from the scope of the LQCD calculations, the prevailing practice
as of now, is to use effective methods to describe the dynamics of the meson 
production reactions. Such methods explicitly include baryon resonance
states, whose properties are extracted by matching predictions of the
theory with the experimental data. The ultimate goal is to compare the
values extracted in this way with those predicted by the LQCD calculations.

In this lecture we review one such method which is based on an effective
Lagrangian model~\cite{shy99}, in the context of $K$ and $\eta$ meson 
(to be referred collectively as $\varphi$ in the following) productions in 
elementary nucleon-nucleon ($NN$) collisions. In this model contributions to
the amplitudes are taken into account by lowest order Feynman diagrams 
(tree-level) which are generated by effective Lagrangians that satisfy
the relevant conservation laws and are consistent with the basic symmetry 
(chiral) of the fundamental theory, namely, the quantum chromodynamics (QCD). 
These Lagrangians involve baryons and mesons as effective degrees of freedom 
instead of quarks and gluons.
 
Recently, there has been a lot of interest in studying the production of
$\varphi$ mesons (which are the next lightest nonstrange members in the
meson mass spectrum) in $NN$ collisions and the corresponding data base has 
enhanced considerably (see, e.g.,~\cite{mos02,sam06,roz06,val07}).
These reactions are expected to provide information on the manifestation of 
the QCD in the non-perturbative regime of energies larger than that of the 
pion physics where concepts like the low energy theorem and partial 
conservation of axial current (PCAC) provide a useful insight into the relevant
physics~\cite{eric88}. The strangeness quantum number introduced by these 
reactions leads to new degrees of freedom into this domain which brings in new
symmetries and conservation laws. They are expected to probe the admixture of
$\bar{s}s$ pairs in the nucleon wave function~\cite{alb96,dov90} and the
baryon-nucleon and $\varphi$-nucleon interactions \cite{del89}.

The elementary $NB\varphi$ ($B$ is a nucleon or hyperon) cross sections are 
one of the most crucial ingredients in the transport model studies of the 
$\varphi$-meson production in the nucleus-nucleus collisions
\cite{mos91,aic91,cas99,cas04}. They provide the opportunity to study the 
equation-of-state of the baryonic matter at high density~\cite{pow02} and 
in-medium modifications of strongly interacting particles which might be 
related to the chiral symmetry restoration \cite{koc97}. These are the 
fundamental aspects of nuclear and hadron physics \cite{sen08}. Furthermore, 
the enhancement of the open strangeness production has been proposed as a 
signature for the formation of quark-gluon plasma in high energy
nucleus-nucleus collisions~\cite{rafe82}. 

The elementary production reactions often provide important means for studying
the properties of the nucleon resonances. For example, the spin-$\frac{1}{2}$,
isospin-$\frac{1}{2}$, and odd parity nucleon resonance $N^*$(1535) 
[$S_{11}(1535)$] has a remarkably large $\eta N$ branching ratio.  It lies very
close to the threshold of the $NN \to NN\eta$ reaction and contributes to the
amplitude of this reaction even at the threshold. Therefore, the study of 
$\eta$ meson production in $NN$ collisions at near threshold beam energies
provides the unique opportunity to investigate the properties of $N^*$(1535) 
which have been the subject of some debate recently (see, e.g.,~\cite{mat05}). 
 
The understanding of the elementary kaon production reactions is a doorway to
the theoretical investigation of the production of hypernuclei in reactions 
like $A(p,K^+){_\Lambda}B$ where the hypernucleus ${_{\Lambda}}B$ has the same
neutron and proton numbers as the target nucleus $A$, with one hyperon added.
Moreover, the attractive nature of the $\eta$-nucleon interaction may lead to 
the formation of bound (quasi-bound) $\eta$-nucleus states (see, e.g., 
\cite{chi91,fix02,hai02,pfe04,sib04}). This subject has been a topic of great 
interest recently. The formation of $\eta$ mesic nuclei depend critically on 
the value of $\eta N$ scattering length ($a_{\eta N}$) (see, e.g.,
\cite{lee08}). The $\eta$ meson production in $NN$ collisions near threshold 
energies provides vital information about $a_{\eta N}$.

\section{{\bf Effective Lagrangian Model of Strange Meson Production in 
Elementary Collisions}}

\begin{figure*}
\begin{tabular}{ccc}
\psfig{file=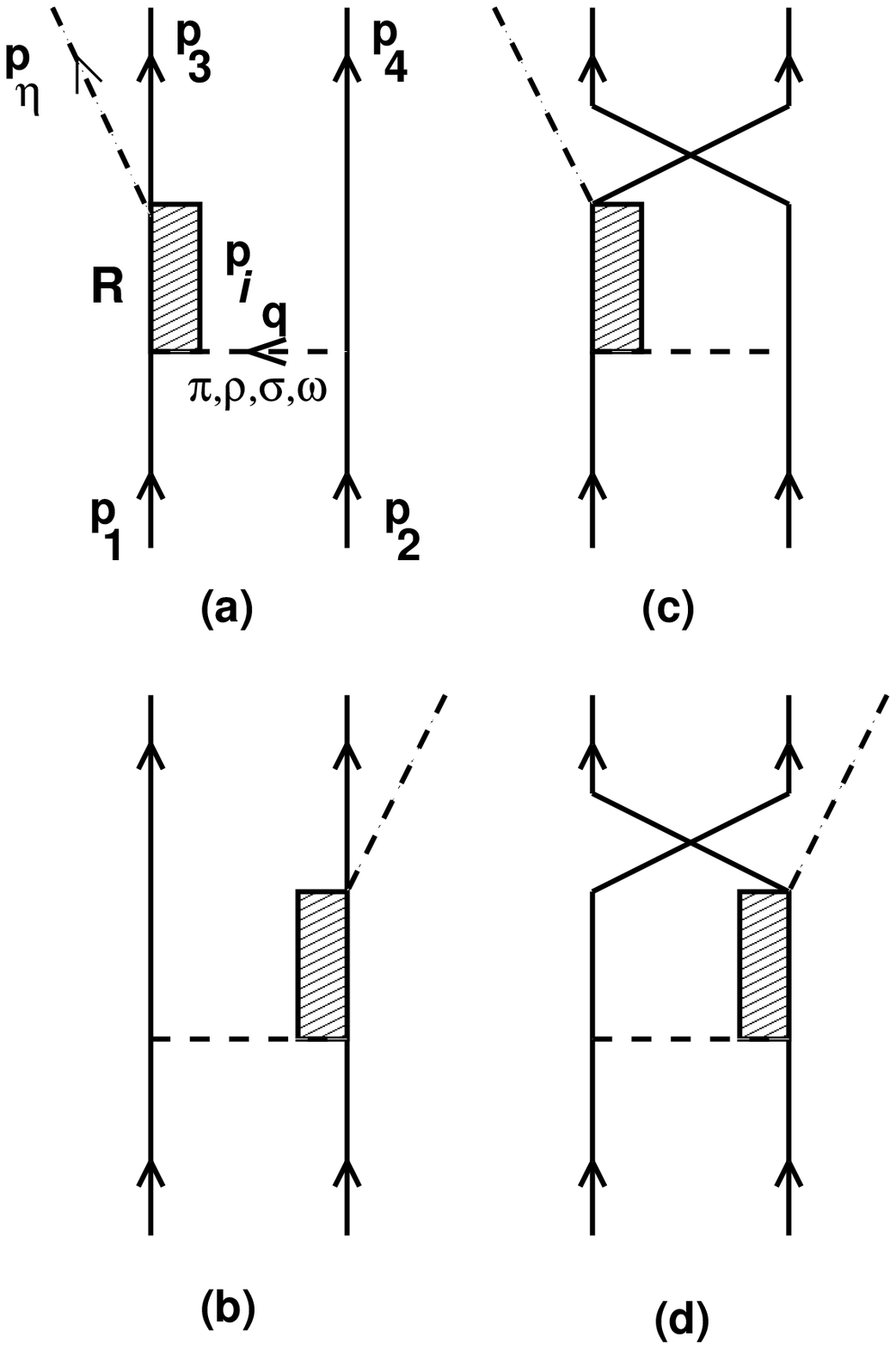,width=5cm} &
\psfig{file=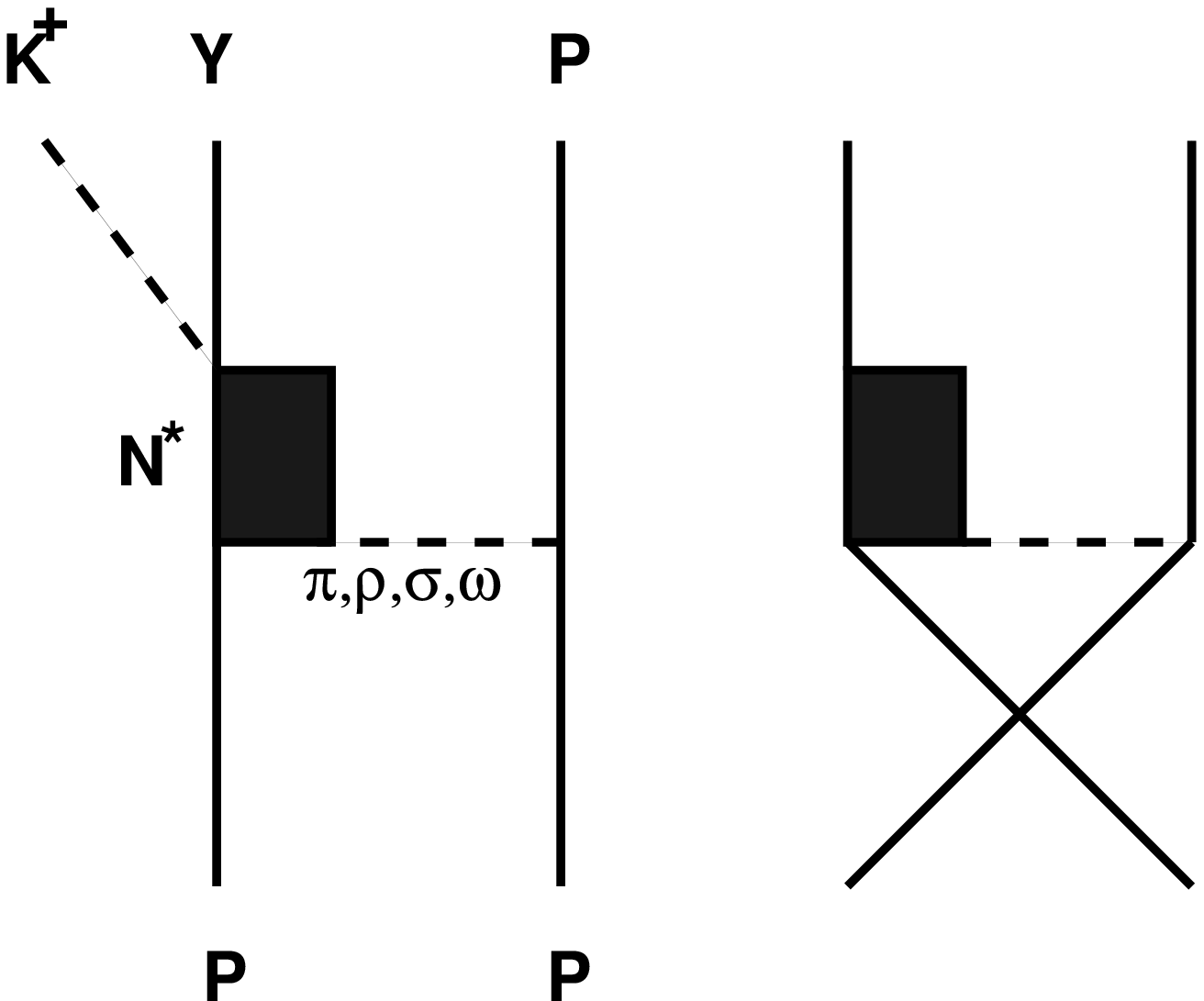,width=6cm}& 
\end{tabular}
\caption{Feynman diagrams for meson production in $NN$ collisions. The graph 
on the left show 4 pieces of the $\eta$ meson production; (a) and (b) are 
the direct target emission and projectile emission processes, while (c)
and (d) are their the exchange counter parts. The graph on the right are the
direct and exchange components of the kaon production.
  }
\end{figure*}

The idea of the effective Lagrangian model is to account for the symmetries 
of the QCD by including only effective degrees of freedom instead of quarks.
These effective degrees of freedom are modeled by baryons and 
mesons which exist as (quasi-)bound quark states. The advantage is that in 
this way one gets a better insight in the underlying production mechanism 
which makes the interpretation of the results easier. However, due to more
complicated interaction structure, the compliance of physical constraints 
like unitarity and analyticity becomes technically more involved. Enforcing
unitarity dynamically requires solving a system of coupled equations with all
possible final states. Recent coupled-channels approaches within an effective 
Lagrangian framework~\cite{lee08,feu98,kor98,pen02,uso05,shk05} are mostly 
confined to reactions leading to two-body final states. None of the effective 
Lagrangian models used to describe reactions leading to three-body final 
states involve in-built unitarity.
 
\subsection{Effective Lagrangians, Form Factors, Coupling Constants}
 
Within our effective Lagrangian approach, we consider the tree-level structure
(Fig.~1) for the amplitudes of meson production in $NN$ collisions. The 
reaction proceeds via excitation of $N^*(1535)$, $N^*(1650)$, and $N^*(1710)$ 
resonances for the $\eta$ meson case and $N^*(1650)$, $N^*(1710)$ and 
$N^*(1720)$ for the kaon case, in the initial interactions between two 
nucleons. These interactions are modeled by the exchange of $\pi$, $\rho$, 
$\omega$ and $\sigma$ mesons. The considered baryonic resonances have 
appreciable branching ratios for decays into the relevant channels. The 
amplitudes are calculated by summations of the Feynman diagrams generated 
by means of the effective Lagrangians at (a) the nucleon-nucleon-intermediate 
meson ($NNM$), (b) resonance-nucleon-intermediate meson ($RNM$), and (c) 
resonance-baryon-final meson ($RB\varphi$) vertices. The assumption entering 
here is that the contributions of higher-order diagrams are either negligible
or they can be absorbed in the form factors of the first order diagrams.

The parameters for $NNM$ vertices are determined by fitting the $NN$ elastic 
scattering T matrix with an effective $NN$ interaction based on $\pi$, $\rho$,
$\omega$ and $\sigma$ meson exchanges. The effective $NNM$ Lagrangians are
\begin{eqnarray}
{\cal L}_{NNM} & = \Big[& -\frac{g_{NN\pi}}{2m_N} {\bar{\Psi}}_N \gamma_5
{\gamma}_{\mu} \tauiso \cdot (\partial ^\mu {\bf \Phi}_\pi) \Psi_N \nonumber \\
 &-& g_{NN\rho} \bar{\Psi}_N \left( \gmu + \frac{k_\rho}
 {2 m_N} \sigma_{\mu\nu} \partial^\nu\right)
 \tauiso \cdot \mbox{\boldmath $\rho$}^\mu \Psi_N  \nonumber \\
 &-& g_{NN\omega} \bar{\Psi}_N \left( \gmu + \frac{k_\omega}
 {2 m_N} \sigma_{\mu\nu} \partial^\nu\right)\omega^\mu \Psi_N \nonumber  \\
 &+& g_{NN\sigma} \bar{\Psi}_N \sigma \Psi_N \Big],
\end{eqnarray}
where $m_N$ denotes the nucleon mass.  Note that Eq.~(1) uses a pseudovector 
(PV) coupling for the $NN\pi$ vertex which is consistent with the chiral
symmetry requirement. Since we use these Lagrangians to directly model the 
$T$-matrix, we have also included a nucleon-nucleon-axial-vector-isovector 
vertex which in the limit of very large axial meson mass, cures the unphysical
behavior in the angular distribution of $NN$ scattering caused by the contact 
term in the one-pion exchange amplitude~\cite{sch94}.  

We introduce, at each interaction vertex, the form factor
\begin{eqnarray}
F_{i}^{NN} & = & \left (\frac{\lambda_i^{2} - m_i^{2}}{\lambda_i^{2} - q_i^{2}}
        \right ), i= \pi, \rho, \sigma, \omega,
\end{eqnarray}
where $q_i$ and $m_i$ are the four momentum and mass of the $i$th exchanged 
meson, respectively. The form factors suppress the contributions of high 
momenta and the parameter $\lambda_i$ which governs the range of suppression,
can be related to the hadron size. Since $NN$ elastic scattering cross sections
decrease gradually with the beam energy (beyond certain value), we take energy
dependent meson-nucleon coupling constants of the following form
\begin{eqnarray}
g(\sqrt{s}) & = & g_{0} exp(-\ell \sqrt{s}),
\end{eqnarray}
in order to reproduce these data in the entire range of beam energies. The
parameters, $g_0$, $\lambda$ and $\ell$ were determined by fitting to the 
elastic proton-proton and proton-neutron scattering data at the beam energies 
in the range of 400 MeV to 4.0 GeV~\cite{sch94}. It may be noted that this 
procedure fixes also the signs of the effective Lagrangians [Eq.~(1)]. The 
values of various parameters are given in Ref.~\cite{shy99}. The same
parameters were used in calculations of all the inelastic channels.

We also require the effective Lagrangians for the $RNM$ and $RB\varphi$
vertices corresponding to all the included resonances. Since the mass of the 
strange quark is much higher than that of the $u-$ or $d-$ quark, one does not
expect the pion like strict chiral constraints for the case of other 
pseudoscalar mesons like $\eta$ and $K$. Thus, one has a choice of pseudoscalar
(PS) or PV couplings for the $NN\varphi$ and $R_{1/2}B\varphi$ vertices
($R_{1/2}$ corresponds to a spin-1/2 resonance). Forms of the corresponding 
effective Lagrangians are given in Ref.~\cite{shy99}. 

The couplings constants for the vertices involving resonances can be determined
from the experimentally observed quantities such as branching ratios for their 
decays to corresponding channels. In most cases however, we have used 
the coupling constants (magnitudes as well their signs) determined in the 
effective Lagrangian coupled-channels analysis of the photon induced meson 
production reactions off nucleon reported in Refs.~\cite{pen02}. Since the 
resonances considered in this study have no known branching ratios for the 
decay into the $N\omega$ channel, we determine the coupling constants for the 
$N^*N\omega$ vertices by the strict vector meson dominance (VMD) hypothesis
which is based essentially on the assumption that the coupling 
of photons on hadrons takes place through a vector meson. In the calculations 
of the amplitudes, propagators are required for intermediate mesons and 
nucleon resonances. These are discussed in Ref.~\cite{shy99}.

\subsection{Final State Interaction}

For describing the data at near threshold beam energies, consideration of
final state interaction (FSI) effects among the three out-going particles is
important. We follow here an approximate scheme in line with the
Watson-Migdal theory of FSI~\cite{wat52}. In this approach the energy
dependence of the cross section due to FSI is separated from that of the
primary production amplitude. This is based on the assumption that the
reaction takes place over a small region of space, a condition fulfilled
rather well in near threshold reactions involving heavy mesons.
The total amplitude is written as
\begin{eqnarray}
A_{fi} & = & M_{fi}(NN \rightarrow NB\varphi) \cdot T_{ff},
\end{eqnarray}
where $M_{fi}(NN \rightarrow NB\varphi)$ is the primary $\varphi$ meson 
production amplitude, and $T_{ff}$ describes the re-scattering among the final
particles which goes to unity in the limit of no FSI. The factorization of the
total amplitude into those of the FSI and primary production, enables one to 
pursue the diagrammatic approach so that the role of various meson exchanges 
and resonances in describing the reaction can still be investigated. 

In those cases where the FSI is confined only to one particular pair of 
particles (mostly among baryons in case of nucleon-baryon-meson final states),
$T_{ff}$ can be calculated by following the Jost function~\cite{wat52} method
(see, e.g., Refs.~\cite{shy99}). If one assumes the relative orbital angular 
momentum between a particular pair of particles to be zero and makes use of 
the effective range expansion of the corresponding phase-shift then the
relevant Jost function can be related to scattering length and effective range
parameters of the scattering between the two particles of that pair. In this 
way, comparison with data at near threshold energies provide a means to get
information about low-energy scattering parameters which are otherwise not 
accessible for unstable particles. 

In certain cases it may be necessary to include FSI among all the three 
outgoing particles since even if the meson-baryon interactions are weak, 
they can still be influential through interference. In the first 
reference cited under~\cite{shy99} expressions are derived for $T_{ff}$ 
in terms of the $T$ matrices describing FSI among all the three two-body 
subsystems of the final state which is obtained by following the technique of
multiple FSI discussed in Ref.~\cite{gil64}.

\section{$NN \to NN\eta$ and $NN \to N\Lambda K$ and $NN \to N\Sigma K$
reactions}
 
After having established the effective Lagrangians, coupling constants
and forms of the propagators, the amplitudes for various diagrams associated
with reactions under study can be calculated in straight forward manner.
The isospin part is treated separately. This gives rise to a constant factor 
for each graph. We emphasize that signs of various amplitudes are fixed by 
those of the effective Lagrangian densities and coupling constants. These 
signs are not allowed to change anywhere in calculations.

\subsection{Check on Vertex Parameters: High Energy Data}

\begin{figure*}
\begin{tabular}{ccc}
\psfig{file=tot_eta-high.eps,width=6cm} &
\psfig{file=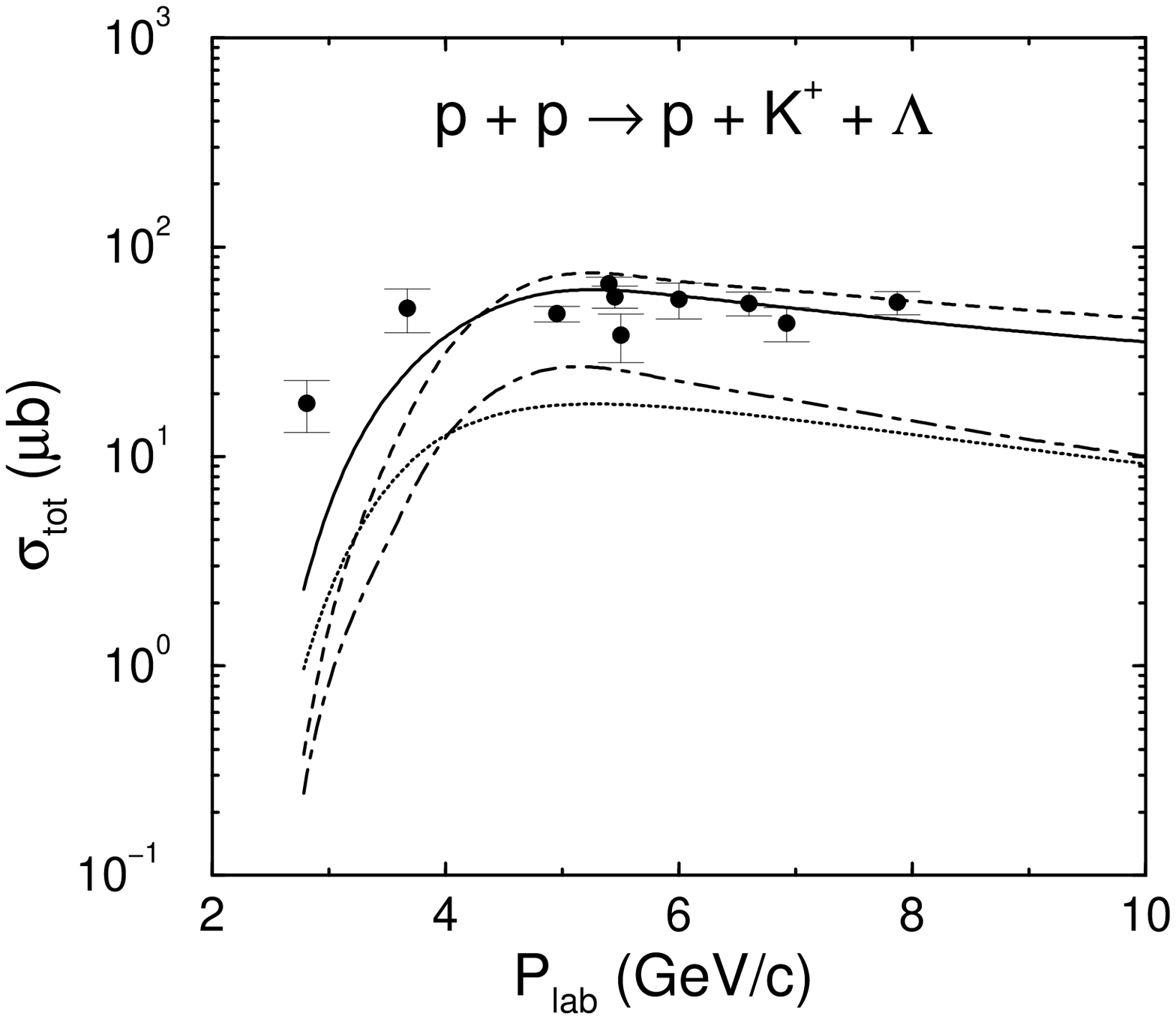,width=6.5cm}& 
\end{tabular}
\caption{
Total cross sections for $p+p \rightarrow p+p+\eta$ (left) and
$p+p \rightarrow p+K^{+}+\Lambda$ (right) reactions as a function of
beam momentum. In the left figure, dashed, dotted and dashed-dotted curves 
represent the contributions of $N^*$(1535), $N^*$(1650), $N^*$(1710) baryonic 
resonance intermediate states, respectively while in the right contributions 
of N$^{*}$(1650), N$^{*}$(1710)and N$^{*}$(1720) resonances are shown by 
dotted, dashed, and dashed-dotted lines, respectively. The coherent sum all 
the resonances is shown by the solid line in each case. The experimental data 
are from Ref.~\protect\cite{lan88}.
}
\end{figure*}
                                                             
In any application of a model, suitability of its input parameters should be
verified on a priority basis. A clean check of the vertex parameters is 
provided by comparison of the calculations with the data at beam momenta 
above 3 GeV/c as FSI effects are expected to be negligible in this region. 
In Fig.~2, we show comparisons of our calculations and the experimental data 
for total cross sections of $pp \to pp\eta$ and $pp \to p\Lambda K^+$ 
reactions. We see that measured cross sections are reproduced reasonably
well by our calculations (solid line) in the entire range of beam momenta. 
This fixes the parameters of all the vertices. In the application of our model
to describe these reactions at near threshold beam energies, the amplitude 
$M_{fi}(NN \rightarrow NB\varphi)$, has been calculated with exactly the same 
values for all the vertex parameters. 

We further notice that while contributions of the $N^*(1535)$ resonance 
dominates $pp\rightarrow pp\eta$ reaction for all the beam momenta, 
the $pp \rightarrow p\Lambda K^+$ reaction is dominated by the $N^*(1710)$ 
and $N^*(1650)$ excitations above and below 3 GeV/c beam momentum, 
respectively. However, in both cases, the interference terms of the amplitudes
corresponding to various resonances are not negligible. 

\subsection{Strangeness production at near threshold beam energies: Role of
Final State Interactions}

Several groups have measured the cross sections for $pp \to pp\eta$,
$pp \to p\Lambda K^+$, $pp \to p\Sigma^0 K^+$, $pp \to n\Sigma^+ K^+$ and
$pp \to p\Sigma^+ K^0$ reactions at beam energies very close to their
respective production thresholds (see, e.g, Ref.~\cite{mos02} for references
upto 2002 and Refs~\cite{roz06,val07} for more recent studies). The data are 
usually presented as functions of kinetic energy in the exit channel. We 
define the excess energy of the reaction as 
$\varepsilon = \sqrt{s}-m_p-m_B-m_\varphi$, where $\sqrt{s}$ is the invariant
mass. At near threshold beam energies, the outgoing particles have
small relative momenta. Therefore, interpretation of the data requires a 
correct treatment of the final state interactions among the final channel
particles.  
\begin{figure*}
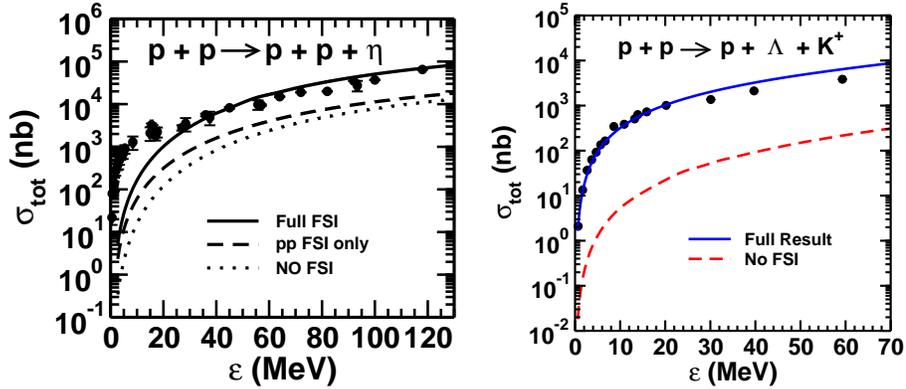

\begin{center}
\begin{tabular}{ccc}
\psfig{file=tot_eta-th.eps,width=6cm} &
\psfig{file=tot_kplam-th.eps,width=5.5cm}& 
\end{tabular}
\end{center}
\caption{
The total cross section for $pp \rightarrow pp\eta$ (left) and 
$pp \to p\Lambda K^+$ (right) reactions as a function of the excess energy.
In the left panel dotted and dashed curves represent cross section obtained 
with FSI effects included only in the proton-proton sub-state of the final 
channel and no FSI at all, respectively. In the right panel results obtained 
with no FSI effects are shown by dashed lines. The solid line shows the 
results obtained with full FSI effects. 
}
\end{figure*}
                                                             
In Fig.~3, we present comparisons of our calculations with the experimental 
data for total cross sections of  $pp \to pp\eta$ and $pp \to p\Lambda K^+$ 
reactions as a function of $\varepsilon$. For both reactions the FSI
effects are important at these low beam energies. While, for the 
$pp \to p\Lambda K^+$ reaction our model is able to describe the data well for
the entire range of beam energies, for the $pp \to pp\eta$ reaction data are 
reproduced only for $\varepsilon$ values in the range of 15 - 130 MeV. However,
for the later case, an important result is that the FSI in the $\eta p$ 
sub-state is indeed quite important in our model. Yet it is not enough to 
explain the data for $\varepsilon <$ 15 MeV. Underpredition of the data by 
theory for these values of $\epsilon$, has also been seen in calculations 
presented in Refs.  \cite{nak02}. 

There may be several reasons for this underprediction. $\eta p$ final state
interaction may have a different form for these low energy which could make it
relatively stronger in this region. We noted that taking larger values for the
scattering lengths $a_{\eta N}$ worsens the fit to the data for $\varepsilon >$
15 MeV. In fact,  best description of the data is provided by the $\eta N$ 
scattering amplitudes corresponding to $a_{\eta N} = 0.51 + i 0.26$. Although,
the real part of this $a_{\eta N}$ is about half of that of the "preferred"
$\eta-N$ parameter set of Ref.~\cite{gre05}, yet a smaller $a_{\eta N}$ is  
consistent with that extracted by several authors
\cite{feu98,gar02,luz02,gas03,lee08} from studies involving meson-nucleon 
scattering. On the other hand, calculations of FSI effects within the 
three-body scattering theory of Faddeev type~\cite{fix04} seem to reproduce 
the data for $\varepsilon <$ 15 Mev. However, predictions of this theory for
$\varepsilon >$ 60 MeV is not known. 

The Born term (nucleon intermediate states) is conjectured to contribute 
strongly at lower energies. To check this point we present in Fig.~4 results
of our full calculations with Born term included. 
\begin{figure*}
\centerline{\psfig{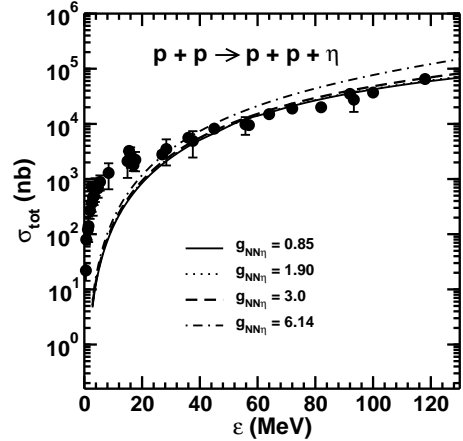}}
\vspace*{2pt} 
\caption{
The total cross section for the $pp \rightarrow pp\eta$ reaction including
the Born term (nucleon intermediate states) as function of $\varepsilon$. 
Results are shown for different values of $g_{NN\eta}$.
}
\end{figure*}
\noindent
We see that the inclusion of the amplitudes corresponding to the nucleon 
intermediate states, makes negligible difference to the results shown in 
Fig.~3 if the value of the coupling constant $g_{NN\eta}$ is below 3.0. 
Using the largest considered value of for $g_{NN\eta}$ (6.14), the results 
are affected to the extent of only a few percent. Obviously, due to 
considerable amount of uncertainty in the value of $g_{NN\eta}$, the nucleon 
excitation amplitudes are quite uncertain and their inclusion makes no 
significant difference in the results reported in Fig.~3. In particular, the 
discrepancy between theory and the data for $\epsilon < 15$ MeV remains 
unresolved.

Recently, it has been shown~\cite{nak08} that the description of the 
$pp \to pp\eta$ data can be improved considerably for lower values of 
$\varepsilon$ within a resonance model by including $D_{13}(1520)$ resonance
with unusually large couplings constants for for $RN\rho$ and $RN\omega$ 
vertices. In this context one should note that coupling constants extracted 
in the effective Lagrangian coupled-channels analysis~\cite{shk05,uso06} of 
$\pi N \to \omega N$ and $\gamma N \to \rho N$ reactions are about an order 
of magnitude smaller than those used in Ref.~\cite{nak08}. We believe that 
explanation of the $pp\eta$ data for beam energies very close to the threshold
is still an open issue. 

The production of light hyperons in proton-proton collisions has been 
extensively studied at close-to-threshold beam energies. The energy dependence
of the experimental total cross sections of $pp \to p\Lambda K^+$ and 
$pp \to p\Sigma^0K^+$ reactions has been well reproduced within the effective 
Lagrangian model (see, last of Ref.~\cite{shy99}). Very recently, data have 
become available also for the $pp \to n\Sigma^+K^+$ reaction. The interest in
the $\Sigma^+$ production channel stems from the fact that it provides a 
sensitive tool to search for a possible penta quark state. COSY-11 and 
COSY-ANKE collaborations have reported quite contrasting values for this 
reaction at excess energies of 13 MeV and 60 MeV~\cite{roz06}, and 129 
MeV~\cite{val07}, respectively. 

In Fig.~5, we compare these data with predictions of our model.
\begin{figure*}
\centerline{\psfig{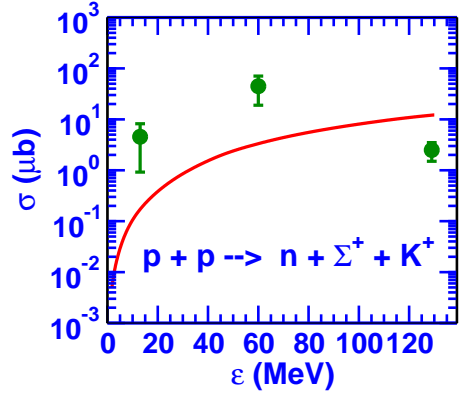}}
\vspace*{2pt}
\caption{
The total cross section for the $pp \rightarrow n\Sigma^+K^+$ reaction as a 
function of excess energy ($\varepsilon$). The experimental data for 
$\varepsilon$ values of 13 MeV and 60 MeV are taken from~\protect\cite{roz06}
and that at 129 MeV is from~\protect\cite{val07}.
}
\end{figure*}
\noindent
We notice that lower energy data points are surprisingly too high - they are  
at least an order of magnitude larger than our calculations. On the other
hand, the higher energy point is smaller than the theoretical results by a 
factor of about 3. Two measurements, thus, imply a very large threshold 
anomaly which must be verified. It has been suggested that inclusion in 
production process of the $\Delta(1620)$ resonance and an unrealistically very
strong $n\Sigma^+$ FSI would allow to achieve a much better agreement (within
factors of 2-4) with the data at lower energies~\cite{xie07}. However,  in this
calculation, the coupling constant for the $\Delta N\Sigma$ vertex is very 
uncertain. While there is no creditable evidence for the decay of 
$\Delta(1620)$ isobar into the $n \Sigma^+$ channel, there are branching
ratios known (albeit with large error bars) for the decay of $\Delta(1600)$
and $\Delta(1920)$ isobars into this channel~\cite{pdg}. One should rather 
include these resonances into the model. In any case, the calculations of Ref.
\cite{xie07} grossly overpredicts the data point at high energy. The solution 
of this problem must await the new measurements of this reaction by the 
COSY-ANKE group at close-to-threshold energies.
  
\section{Summary and Conclusions}

In summary, we presented some application of our effective Lagrangian model
in understanding the recent data on hidden strangeness production 
($NN \to NN\eta$) and open strangeness production ($pp \to p\Lambda K^+$,
$pp \to p\Sigma^0K^+$, $pp \to n\Sigma^+ K^+$) reactions in nucleon-nucleon
collisions.  With the same set of vertex parameters, the model is able to 
provide a good description of the $pp \to pp \eta$ reaction at higher 
as well as near threshold beam energies except for the excess energies below
15 MeV where our calculations underpredict the experimental cross sections. 
We examined the validity of several suggestions for understanding this 
discrepancy. It turns out that a very strong $\eta N$ FSI or the 
inclusion of the Born term is unlikely to explain this anomaly. Inclusion 
of the $D_{13}(1520)$ resonance with reasonable coupling constants at the 
vector meson vertices is also unlikely to be of any help. This at the moment 
remains an open issue.

This model is able to describe  majority of the data on $pp \to p\Lambda K^+$
and $pp \to p\Sigma^0K^+$ reactions where it is found that the $N^*$(1650) 
resonant state contributes predominantly to both these reactions at near 
threshold beam energies. Therefore, the study of these reactions provides an 
ideal means of investigating the properties of this $S_{11}$ baryonic 
resonance. On the other hand, calculations are not compatible with the 
published few data points on $pp \to n\Sigma^+K^+$ reaction. One has to wait 
for the results of the new measurements for this reaction by the COSY-ANKE 
group at near threshold beam energies.

\section{Acknowledgments}

The author wishes the thank Hans Str\"oher for several useful conversations 
on the measurements of the COSY-ANKE group.

\end{document}